\documentclass[aps,prb,twocolumn,groupedaddress,floatfix,showpacs]{revtex4}

\usepackage{amsmath}
\usepackage{graphicx}
\usepackage{epsfig}

\begin{document}

\title{Carrier relaxation in GaAs v-groove quantum wires and the effects of localization}

\author{N.\ I.\ Cade}
\email{n.cade@physics.ox.ac.uk}
\author{R.\ Roshan}
\author{M.\ Hauert}
\author{A.\ C.\ Maciel}
\author{J.\ F.\ Ryan}
\affiliation{Clarendon Laboratory, University of Oxford, Oxford, OX1 3PU, UK}

\author{A.\ Schwarz}
\author{Th.\ Sch\"{a}pers}
\author{H.\ L\"{u}th}
\affiliation{Institut f\"{u}r Schichten und Grenzfl\"{a}chen, Forschungszentrum, 52425 J\"{u}lich, Germany}


\begin{abstract}
Carrier relaxation processes have been investigated in GaAs/AlGaAs v-groove quantum wires (QWRs) with a large
subband separation ($\Delta E\simeq46$ meV). Signatures of inhibited carrier relaxation mechanisms are seen in
temperature-dependent photoluminescence (PL) and photoluminescence-excitation (PLE) measurements; we observe
strong emission from the first excited state of the QWR below $\sim$50 K. This is attributed to reduced inter-subband relaxation via phonon scattering between localized states. Theoretical calculations and experimental results indicate that the pinch-off regions, which
provide additional two-dimensional confinement for the QWR structure, have a blocking effect on relaxation
mechanisms for certain structures within the v-groove. Time-resolved PL measurements show that
efficient carrier relaxation from excited QWR states into the ground state, occurs only at temperatures $\agt$ 30
K. Values for the low temperature radiative lifetimes of the ground- and first excited-state excitons have been
obtained (340 ps and 160 ps respectively), and their corresponding localization lengths along the wire estimated.
\end{abstract}

\pacs{78.67.Lt, 73.21.Hb, 78.47.+p}

\maketitle
\section{Introduction}
\label{intro}

In recent years there has been considerable interest in semiconductor v-groove structures, as a means of providing
a high quality quasi-one-dimensional (1D) electronic system. This has enabled the study of fundamental physical
properties in confined systems, as well as the development of novel optoelectronic structures such as quantum wire
lasers.\cite{sirigu} Strong spatial confinement and large subband separation are essential features for room
temperature devices using these structures. However, during the growth of low-dimensional structures, interfacial
roughness develops due to misorientation with the substrate,\cite{reinhardt} and monolayer (ML)
fluctuations.\cite{lelarge} This results in the localization of excitons at low temperatures.\cite{bastard,D_liu}
In 1D wires these width fluctuations lead to the formation of quasi-0D quantum boxes along the wire
axis,\cite{hasen} the effects of which have been observed in $\mu$PL
measurements\cite{bellessa,vouilloz,tribe1998} and exciton radiative lifetimes.\cite{oberli1999} Localized
excitons have a key role in producing population inversion for laser emission,\cite{sirigu} and strongly affect
the transport characteristics of 1D structures.\cite{nic}

In this report we present time-integrated and time-resolved PL measurements, and PL-excitation measurements, on an
array of GaAs QWRs with a subband separation $>k_{B}T_{\text{room}}$.  These measurements reveal a strong
temperature dependence of carrier relaxation mechanisms between higher excited states into the ground state of the
quantum wires. We find evidence of a non-thermal exciton distribution at low temperature, and ascribe this to
strong exciton localization. These effects are also evident in temperature-dependent measurements of excitonic
lifetimes.

\subsection{Structure}

\begin{figure}[tb]
\epsfig{file=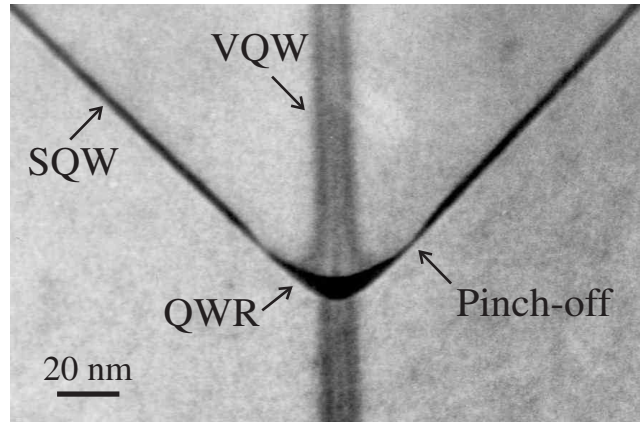,width=8.4cm}
 \caption{\label{tem} TEM micrograph cross-section showing the crescent-shaped
QWR, which is separated from side QWs by  narrow pinch-off regions. The Ga rich VQW that also forms is seen to
consist of three separate structures.}
\end{figure}

The QWR structure studied here was prepared using a semi-insulating (100) GaAs substrate patterned with an array
of [01$\overline{1}$]-oriented 3 $\mu$m pitch v-grooves; each groove was separated by a Si0$_{2}$ photolithographic mask. A nominal 3 nm GaAs quantum well (QW) was grown within
undoped Al$_{0.3}$Ga$_{0.7}$As barriers by low-pressure metal-organic chemical-vapor deposition (MOCVD). Polycrystalline growth occurs on the residual Si0$_{2}$ between the grooves, which prevents the formation of a planar (100) QW at the top of the side-wall quantum wells (SQWs).\cite{schapers} A cross-sectional
view of a similar heterostructure is shown in the transmission electron microscope (TEM) image of Fig.\ \ref{tem}. A
crescent-shaped GaAs QWR forms at the vertex of two SQWs, separated from them by a
narrow constriction or pinch-off. The well thickness in these narrow regions is 1.4 nm,  and the vertical
thickness of the QWR is 7.6 nm, resulting in strong 2D confinement in the wire. A vertical quantum well (VQW) also
forms at the center of the groove, perpendicular to the substrate, due to the higher mobility of Ga atoms in the
AlGaAs.\cite{biasiol96} More detailed reports on the growth and characterization of similar grooves are given
elsewhere.\cite{schapers,cade,marc}

Trapping of carriers into the QWR, and the relaxation dynamics of excited states, are important issues when
considering carrier injection into a 1D structure. In QWR lasers carriers entering the wire region must be able
to scatter into the appropriate low energy lasing states by processes such as optical phonon emission and
carrier-carrier scattering. In a v-groove structure there are two main carrier capture mechanisms: direct capture
from the barrier into the QWR (3D - 1D), or capture from the barrier into the SQW and VQW and then subsequent
diffusion and trapping into the QWR (3D - 2D - 1D).\cite{walther,kan} Calculations by Kiener \textit{et
al.}\cite{kiener} of longitudinal-optical (LO) phonon scattering rates in a v-groove structure, show that the
first process is greatly reduced compared to barrier - VQW - QWR scattering, due to the small spatial overlap
between initial and final states and their relatively large energy separation. Composition nonuniformity in the
barrier may also significantly affect carrier capture efficiency.\cite{wang95} Capture into the QWR from the SQW
is complicated due to the formation of the pinch-off regions, and will be discussed in this report.

\subsection{Electronic states}
\label{electronic}

\begin{figure*}[tb]
\includegraphics[width=12.9cm]{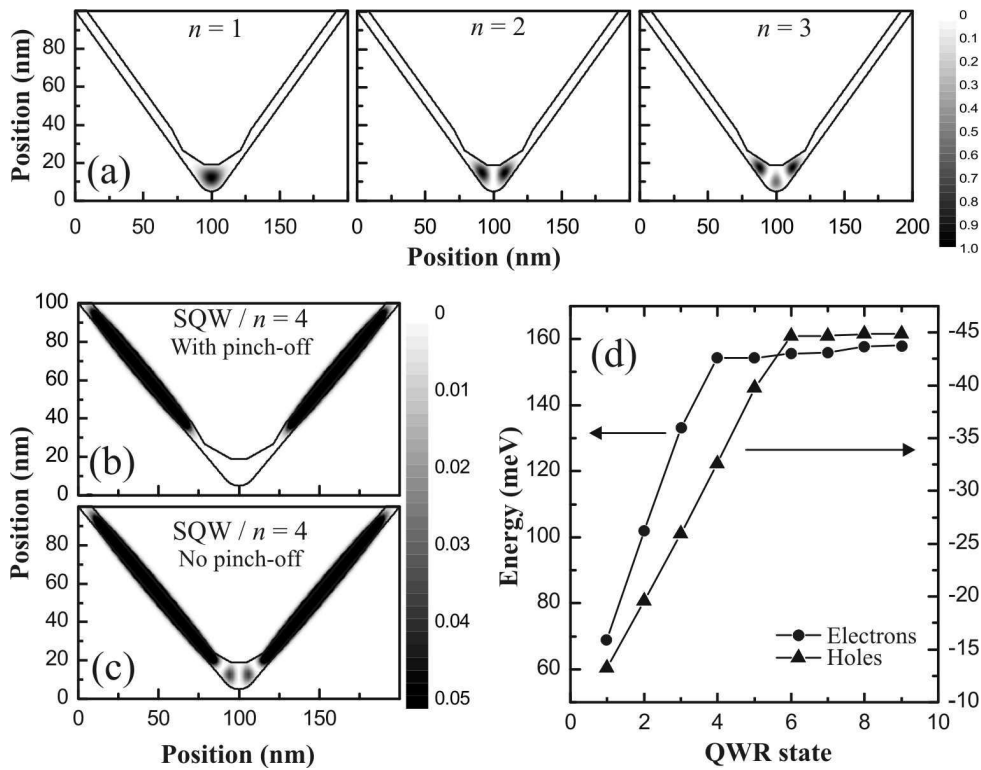}
 \caption{\label{theory} (a) First three QWR electronic states (with pinch-off);
(b) SQW / $n=4$ state with pinch-off; (c) as (b) without pinch-off barrier included in the calculation. The shaded
scales indicate the normalized relative probability density. (d) Calculated electron (circle) and heavy-hole
(triangle) confinement energies for the sample under investigation.}
\end{figure*}

Electronic states of the quantum wire have been determined using a plane wave expansion to solve
the Schr\"{o}dinger equation for a real QWR cross-section.\cite{Eoin,axel}
These QWR profiles have been obtained directly from TEM images such as Fig.\ \ref{tem}, and used as input data.
However, in order to avoid the influences of varying image quality, the profiles shown in Fig.\ \ref{theory} are
based on a systematic extraction of facet growth evolution.\cite{biasiol98} The size of the real-space window was
limited by computational considerations, which meant that the full SQW length could not be included. The quasi-2D
side-well states are therefore seen as higher energy states of the QWR ($n\geq4$ for electrons).

Figure \ref{theory}(a) shows the electron probability density functions for the first three confined QWR states.
Calculations for a 6 nm QW thickness are shown for clarity, but the results are qualitatively the same for a 3 nm
QW. All three electron states are confined within the crescent-shaped wire region, but the first two are localized
on different facets. More specifically, the $n=2$ QWR state has a high probability density over the \{311\}A
facets which form near the bottom of the groove. These facets  are expected to have quasi-periodic modulations due
to step bunching,\cite{biasiol97} which will result in strong localization effects for excitons in the $n=2$
state.

Figures \ref{theory}(b) and (c) show the $n=4$ state obtained with and without a pinch-off region respectively. In the latter case the constriction is removed numerically from the QWR structural profile. This pinch-off produces an additional
potential barrier between the quasi-1D and 2D states, strongly restricting $n=4$ electrons to the SQWs,
with vanishing probability in the QWR itself. This results in a reduction of the overlap between QWR and SQW
electronic wavefunctions, and hence effects the efficiency of carrier relaxation from the SQW into the QWR.  Fig.\
\ref{theory}(d) shows the calculated confinement energies of the electron and heavy-hole states for the specific
sample under investigation. A transition from confined QWR states to quasi-2D side-well states can been seen at
$n=4$ and $n=6$ for the electron and hole states respectively. These calculations have been performed using
parabolic band approximations, whereas in reality the valence band-structure is non-parabolic due to heavy- and
light-hole mixing.\cite{sercel,bockelmann92} The effects of the latter can be ignored if the light-hole confinement is
large compared to typical heavy-hole subband separations. The heavy- / light-hole character also changes for the
different subbands: calculations based on samples similar to the one considered here, predict the first valence
subband $h_{1}$ to be $\sim$90\% heavy-hole-like but $h_{6}$ is 70\% light-hole-like.\cite{vouilloz_holes}

\section{Results and discussion}
\label{results}

PL measurements were made using either an Ar$^{+}$ laser operating at 2.41 eV, or a tuneable dye laser ($\sim $1.6
- 1.8 eV),  in order to excite above or below the AlGaAs band-gap. The dye laser was also used for PLE
measurements. Luminescence was detected with a double monochromator and either a Peltier cooled PMT or nitrogen
cooled CCD depending on the experimental setup. Time-resolved PL measurements were performed using a mode-locked
Ti:Sapphire laser and a synchronous streak camera, with an overall time resolution of $\sim$20 ps.

\subsection{Photoluminescence measurements}
\label{pl}

\begin{figure}[tb]
\epsfig{file=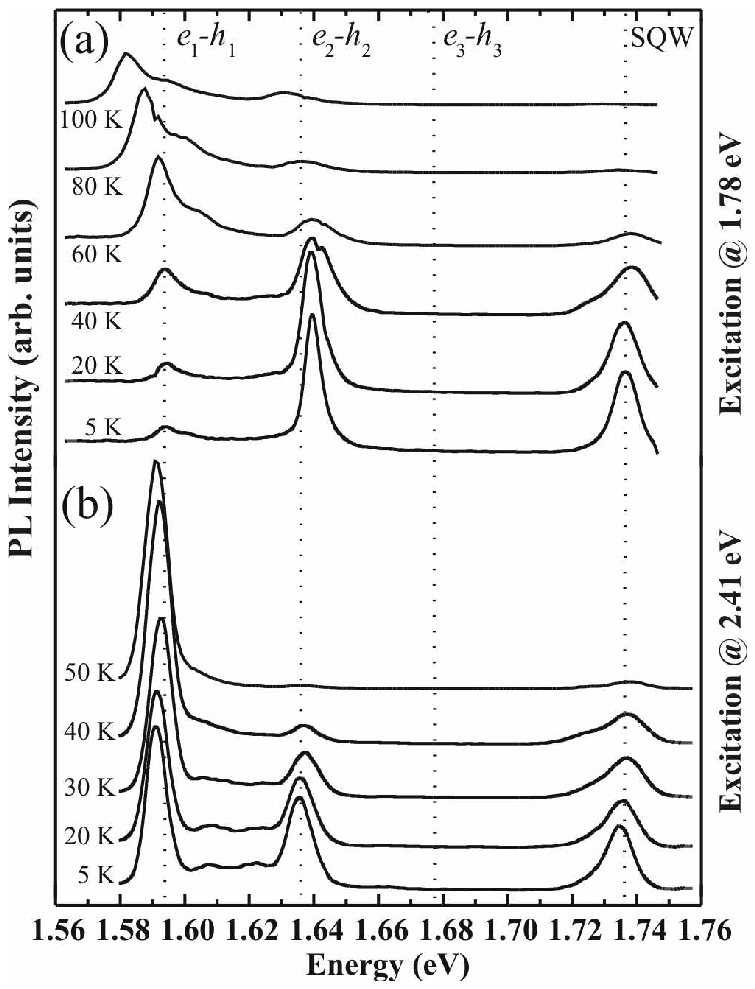,width=8.4cm} \caption{\label{118tempPL} Temperature-dependent PL spectra, exciting at
(a) 1.78 eV  and (b) 2.41 eV.  The dotted lines give the theoretical energies for various QWR optical
transitions.}
\end{figure}

Figure \ref{118tempPL} shows the v-groove luminescence spectra as a function of temperature, when exciting with
(a) the dye laser at 1.78 eV and (b) the Ar$^{+}$ laser at 2.41 eV. In the former case, only the QWR and SQW structures are excited, whereas with the Ar laser all structures and the AlGaAs barrier are excited. The calculated optical transitions from the QWR are also
shown. From now on we will write the
electron-hole transition $e_{x}-h_{x}$ as $n=x$. All of the calculated transition energies have been rigidly
shifted so that the $n=1$ line corresponds to the low temperature peak at 1.593 eV in Fig.\ \ref{118tempPL}(a),
which is attributed to excitonic recombination from the $n=1$ state. The peak at 1.639 eV corresponds closely to
the calculated value for recombination from the $n=2$ state.  These assignments have been confirmed by magneto-PL measurements\cite{angela} and scanning near-field optical microscopy (SNOM);\cite{marc} doped samples with otherwise identical growth sequences also exhibit an enhanced Fermi Edge Singularity (FES) due to the proximity of the $n=2$ state to the Fermi level.\cite{angela,rodriguez}

The line-widths $\delta$$E$$_{\text{PL}}$ of the $n=1$ and $n=2$ QWR peaks are approximately 6 meV and 7 meV
respectively at 5K; SNOM measurements on individual wires indicate that this broadening is due to intra-wire inhomogeneities. The energy separation between the first two subband transitions $\Delta$$E$ is 46 meV; this large subband separation results from the small QW width and relatively strong pinch off, as well as the specific MOCVD growth parameters.\cite{otterburg} The ratio $\rho_{\text{PL}} \equiv \Delta$$E$$/$$\delta$$E$$_{\text{PL}} \simeq$ 7.7, which is comparable to the best values of  $\rho_{\text{PL}}$ that have recently been achieved.\cite{otterburg,wang1}

In both sets of measurements we find that there is a decrease in the $n=2$ luminescence intensity with increasing
temperature.  Simultaneously the $n=1$ intensity increases, indicating that inter-subband relaxation of carriers
is enhanced at higher temperatures due to scattering between localized states. This will be discussed in more
detail with the results of the PLE measurements described below.

A predicted $n=3$ state at 1.675 eV does not appear in the PL spectra, but is clearly evident in the PLE measurements. This indicates efficient relaxation to the $n=1$ state due to a significant wavefunction overlap, as seen in Fig.\ \ref{theory}(a). Luminescence from the side- and vertical QWs is detected at 1.735 eV and 1.804 eV respectively; in the former case, this corresponds to a well thickness of approximately 2.2 nm.

The power density for both excitation energies was 50 W$\,$cm$^{-2}$; similar effects are observed down to 50 mW$\,$cm$^{-2}$ corresponding to a photo-excited carrier density of $\sim$10$^2$ cm$^{-1}$. When using the Ar laser, absorption
in a large volume of the  barrier and subsequent carrier diffusion into the QWR results in a much greater
($\sim$50$\times$) wire carrier density than for excitation with the dye laser at 1.78 eV. For the latter case,
the carrier population in the QWR is cooled efficiently by phonon emission; with Ar
illumination absorption in the barrier generates a hot-phonon population that will facilitate exciton detrapping in the QWR.
Direct capture from the VQW to the $n=1$ state is also possible with Ar illumination. This may explain why the ratio of the intensities of the $n=1$ and $n=2$ peaks is different for the two excitation
energies: with Ar illumination the $n=2$ peak vanishes at 50 K, whereas it is still clearly observed at 100 K when
exciting at 1.78 eV.

The low temperature spectra in Fig.\ \ref{118tempPL}(b) show small luminescence peaks at 1.606 eV and 1.622 eV,
between the $n=1$ and $n=2$ peaks. Previous magneto-PL measurements on similar samples indicate that these may
originate from higher valence band $e_{1}-h_{n}$ excitonic transitions.\cite{angela}  Parity violating transitions are optically forbidden by selection rules for an ideal QWR,\cite{bockelmann92} but are allowed for an asymmetric wire potential, which is likely to be the case for real
QWRs. Figure \ref{118tempPL}(a) also shows that, when exciting below the AlGaAs band-gap,  the $n = 1$  peak develops a high
energy shoulder above 40 K; the temperature-dependence of this shoulder reflects the increased thermal population of the higher valence subbands. The red-shift observed in all of the peaks above 40 K is due to the overall reduction of the band-gap.

The temperature dependence of the SQW luminescence peak provides information on the real-space transfer dynamics
of the structure. At 5 K there is strong luminescence with both excitation sources, which decreases with
increasing temperature: the peak has almost disappeared by 50 K for excitation at 2.41 eV, but persists until
$\sim$80 K when excited at 1.78 eV.

Two types of potential barrier may exist that can inhibit relaxation between the SQW and QWR. Monolayer
fluctuations in the \{111\} side QW will trap excitons in potential minima, and can have a detrapping time longer
than the recombination time at low temperatures.\cite{bastard} With increasing temperature, excitons will be
thermally activated and subsequently captured into the QWR. This mechanism results in a blue-shift in the low
temperature SQW PL peak position with increasing temperature. As mentioned in Sec.\ \ref{electronic}, there is also
a potential barrier due to the pinch-off regions, which provide 2D confinement within the quantum
wire.\cite{liu} Adiabatic calculations for this sample indicate that the pinch-off barrier is $\sim$75 meV and 40
meV for electrons and heavy-holes respectively, which is consistent with the transition between confined QWR states and
SQW states seen in Fig.\ \ref{theory}(d). The spectra in Fig.\ \ref{118tempPL} do show a small ($\sim$3 meV)
blue-shift in the SQW peak position over the range 5 K - 50 K, indicating that there is some activation out of
potential minima. However, the intensity of the SQW peak only starts to decrease significantly at 40 K, above which it
diminishes rapidly with increasing temperature. This indicates that the pinch-off regions produce the dominant barrier for
carrier transfer into the QWR.

For all of the PL peaks observed in Fig.\ \ref{118tempPL}, the magnitude of any thermally  induced energy shift is
much less than the sum of the exciton binding energy ($\sim$10 meV)\cite{jkim} and Stokes shift at 5 K. This
indicates that  emission from the QWR is dominated by excitonic recombination over the whole temperature range
measured.\cite{oberli97}

\subsection{Photoluminescence excitation measurements}
\label{plesection}

\begin{figure}[tb]
\epsfig{file=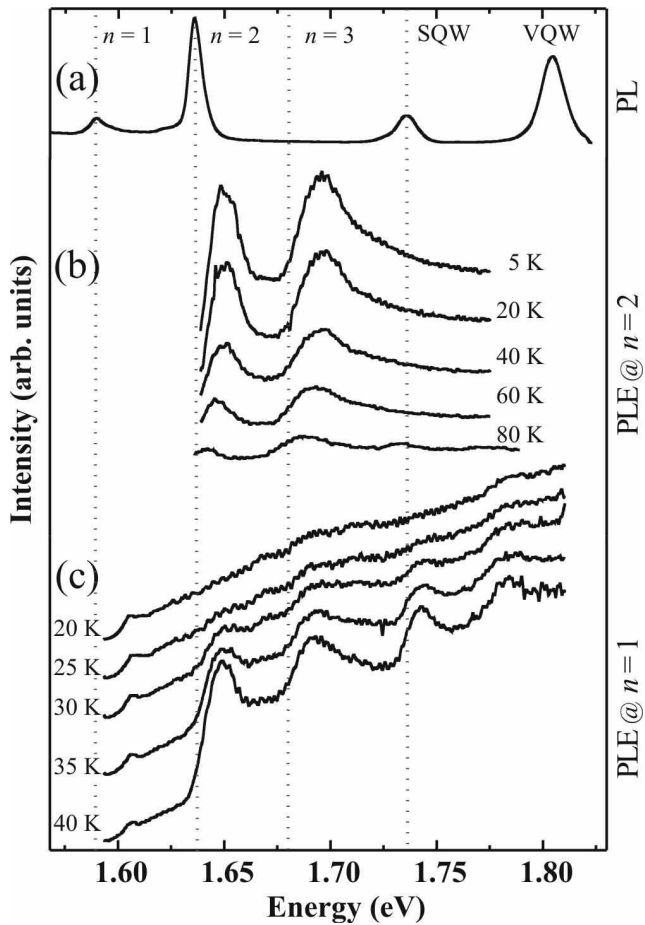,width=8.4cm} \caption{\label{118ple1} (a) PL spectrum at 5 K, exciting at 1.83 eV. (b) PLE
spectra as a function of temperature,  detecting at the $n=2$ QWR peak energy, and (c) detecting at the $n=1$ QWR
peak energy.}
\end{figure}

Figure \ref{118ple1} shows the PL spectrum taken at 5 K, and temperature-dependent PLE spectra obtained while
detecting at  the $n=1$ and 2 transition peaks. The PLE spectra for the $n=2$ luminescence, in Fig.\
\ref{118ple1}(b), show clear absorption from the $n=2$ transition with a Stokes shift of 11 meV, as discussed
below. The peak at 1.695 eV is assigned to absorption from the $n=3$ transition, which is in good agreement with
the calculated transition energy. The width of this PLE peak is greater than that of the $n = 2$ peak, which is
consistent with higher excited states being more sensitive to fluctuations in the confinement potential. With an
increase in temperature, the intensities of both peaks decrease as fast inter-subband relaxation to the ground
state occurs. Above 60 K the absorption peaks shift to lower energies due to the reduction of the band-gap.

When the detection is set at the $n=1$ PL peak, in Fig.\ \ref{118ple1}(c), a weak absorption peak is observed at
1.61 eV, but the spectrum is otherwise featureless below 20 K. This is consistent with the behavior of incomplete
relaxation from higher energy excited states to the QWR ground state at low temperatures.  The absorption at 1.61 eV is unlikely to be from the $e_{1}-h_{1}$ QWR transition as this would give a Stokes shift of almost 20 meV, much larger than for the $n = 2$ transition;\cite{wang} it is most probably from a parity violating transition (e.g.\ $e_{1}-h_{2}$). In order to measure the $n=1$ PLE response it is necessary to set the detection to a low energy tail state of the PL peak; however the QWR luminescence is not strong enough to do this.

By 40 K strong features corresponding to $n=2$ and $n=3$ absorption appear in the spectra; absorption is also observed from
$e_{1}-hh_{1}$ and $e_{1}-lh_{1}$ transitions in the SQW, at 1.74 eV and 1.785 eV respectively. The transition
between the regime of incomplete and efficient relaxation from higher states is clearly seen to occur over a small
temperature range at $\sim$30 K. The strong SQW absorption peaks appearing in the PLE also indicate efficient
carrier transfer from the SQW into the QWR above 30 K, as discussed for Fig.\ \ref{118tempPL}. Interestingly,
absorption from the SQW is not seen below 80 K when detecting at the $n=2$ energy in Fig.\ \ref{118ple1}(b). This
indicates that once carriers are able to transfer through the pinch-off region, they will preferentially recombine
in the ground state of the QWR.

There are a number of mechanisms that can explain the inhibited inter-subband relaxation observed at low
temperatures:

Figure \ref{tem} shows that the VQW consists of three separate Ga rich regions that extend upwards from different
facets of the QWR. These sub-structures of the VQW have an Al content of $\sim$22\% compared to 30\% in the
barrier,\cite{axel} thus generating a laterally varying potential.\cite{biasiol96} The first two QWR states have
a real-space separation over the wire region, as shown in Fig.\ \ref{theory}, thus the lateral
potential may be sufficient to restrict relaxation from the $n=2$ QWR state at low temperature.

The sudden transition in the temperature-dependent PL and PLE spectra can also be explained by considering the role of
phonons in carrier relaxation mechanisms. Generally, photo-generated electron-hole pairs will rapidly form excitons within $\sim 20$ ps as they simultaneously
relax to the band minima:\cite{damen90} In polar semiconductors, a hot carrier distribution can cool efficiently
by emission of LO-phonons of energy $\hbar\omega_{\text{LO}}$ (36 meV for GaAs), with a characteristic time of
$\sim$10$^{-13}$ s. However, in 1- and 0-D structures at low temperatures, the emission of high energy LO-phonons
may be strongly repressed due to energy and momentum conservation,\cite{bockelmann90,zheng} resulting in the
controversial phonon 'bottle-neck' effect.\cite{bensity91} In quantum wires this is usually manifested as a
reduction in inter-subband relaxation.\cite{grundmann94} Longitudinal-acoustic (LA) phonons are expected to give
the dominant contribution to low energy dissipation processes ($<$$\hbar\omega_{\text{LO}}$), but with a
relatively much longer scattering time ($\sim$10$^{-9}$ s). In our case, we find that the electron subband
separation is comparable to $\hbar\omega_{\text{LO}}$. Even if the first-order LO-phonon coupling is forbidden,
LO$\pm$LA multiphonon processes should create a window of rapid (subnanosecond) relaxation around
$\hbar\omega_{\text{LO}}$.\cite{inoshita92} However, the low temperature relaxation dynamics of photoexcited
carriers are also expected to be strongly influenced by disorder. At low temperature, excitons in quasi-0D quantum
dots exhibit a strong decrease in exciton-phonon scattering rates with increasing spatial
quantization.\cite{bockelmann93} For strong confinement, relaxation by LA-phonon emission is much less efficient
than radiative recombination, leading to very weak ground-state luminescence.

For the QWR system considered here, the $n=2$ subband is subjected to large interfacial disorder, and may be
considered as a series of localized 0D states with a 1D continuum of delocalized states. By considering energy and
momentum conservation arguments, the maximum acoustic phonon energy that results  in significant scattering rates
between quasi-0D quantum boxes is given by $E_{\text{max}}=hc_{s}/L_{z}$,\cite{bensity95} where $c_{s}$ is the
sound velocity in the material (5150 ms$^{-1}$ in GaAs) and $L_{z}$ is the dimension of the strongest confinement.
This latter value is 7.6 nm for our quantum wires, giving a maximum acoustic phonon energy of about 2.8 meV. Hence
we would expect the peak for this phonon population to occur at a lattice temperature of $\simeq$ 30 K. Below this
temperature, exciton relaxation will be effectively inhibited due to suppression of acoustic phonon emission over
the energy range of the localized states.\cite{hasen,oberli2000} Excitons will then recombine radiatively, resulting in
strong luminescence at the $n=2$ level. At higher temperatures the localized excitons can relax into the ground state via phonon-assisted real-space scattering,\cite{bastard} within the exciton recombination time. This scenario is consistent with the sudden transition observed in the PLE spectra of Fig.\ \ref{118ple1}(c) between 30 - 40 K.

\begin{figure}[tb]
\epsfig{file=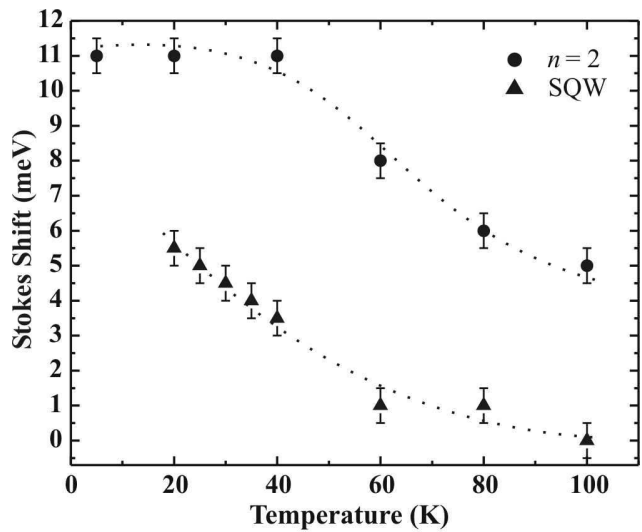,width=8.4cm} \caption{\label{118stokes} Temperature dependence of the Stokes shift for the
$n=2$ (circles) and SQW (triangles) transitions. The dotted lines are guides to the eye.}
\end{figure}

Figure \ref{118stokes} shows the temperature dependence of the Stokes shift for the $n=2$ and SQW transitions,
obtained from the PL and PLE spectra. In both cases the decrease of the Stokes shift with temperature reflects the
thermal redistribution of excitons from localized states to extended states. For the SQW, the Stokes shift drops
to almost zero at around 60 K, which indicates that the majority of the exciton population has become delocalized.
The Stokes shift for the $n=2$ transition is relatively high at low temperatures,\cite{wang} and only starts to
decrease around 40 K. The finite Stokes shift at 100 K indicates that there are still a significant number of $n=2$ excitons in localized states at this temperature, which is consistent with the persistence of the PL in Fig.\  \ref{118tempPL}.

\subsection{Time-resolved measurements}

\begin{figure}[tb]
\epsfig{file=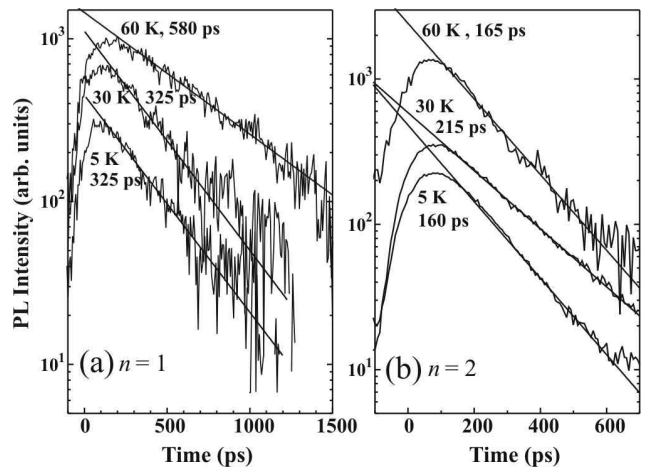,width=8.4cm} \caption{\label{118decay} Temporal profiles of the (a) $n=1$ and (b) $n=2$
luminescence peaks at three different temperatures. The monoexponential fits used to extract the decay times are
shown.}
\end{figure}

\begin{figure*}[tb]
\epsfig{file=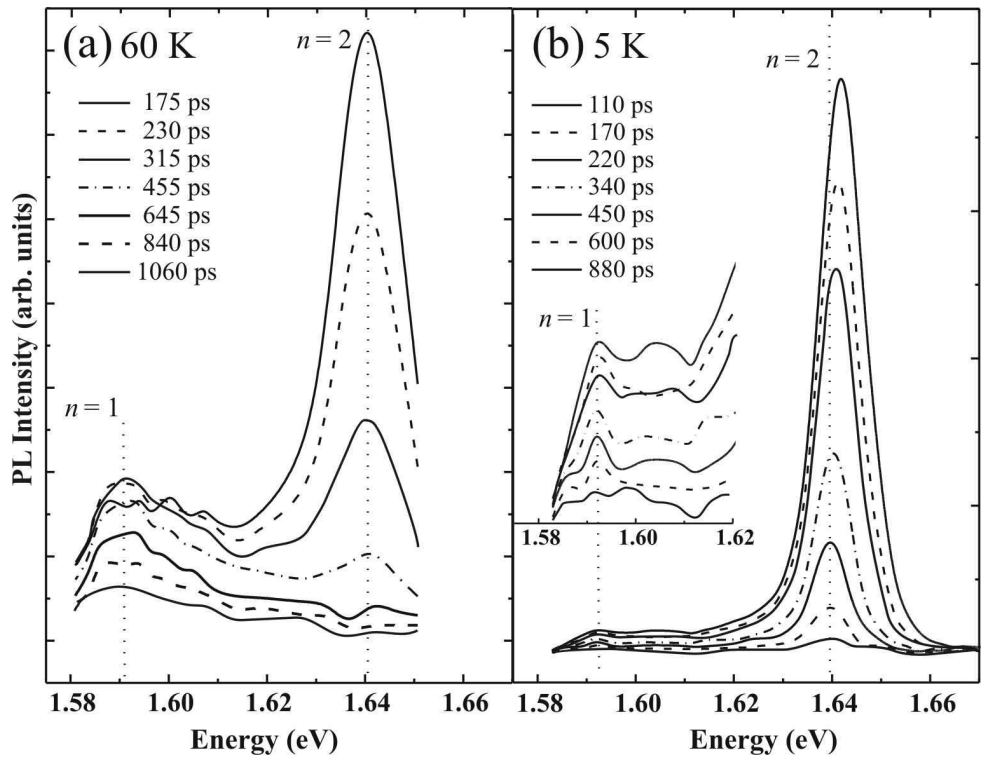,width=12.9cm} \caption{\label{pldecay} Transient PL spectra (smoothed) at (a) 60 K and (b) 5
K. The inset in (b) is a magnified view of the $n=1$ peak. Note the shift in the $n=2$ peak in (b).}
\end{figure*}

Time-resolved PL measurements were made by exciting the sample with a mode-locked Ti:Sapphire laser, operating at
76 MHz and 1.74 eV. Figures \ref{118decay}(a) and (b) show the temporal profiles of the $n=1$ and $n=2$ PL peaks
respectively, at three different lattice temperatures. All of the measured profiles show a monoexponential decay
of the luminescence of the form $\exp(-t/\tau)$, over at least one order of magnitude,  from which the decay time
$\tau$ is obtained. By measuring the decay profiles for a range of detection energies, the temporal evolution of
the PL spectra  were obtained and are shown in Fig.\ \ref{pldecay} at (a) 60 K and (b) 5 K. Due to the limited set of detection energies sampled, it is difficult to speculate on the existence of other peaks close to the QWR ground state. At 60 K, the $n=2$
peak drops very quickly and has vanished by $\sim$600 ps. However, the $n=1$ peak intensity remains essentially
unchanged up to $\sim$450 ps, and the emission persists at the longest measurement time. This is due to
repopulation of the ground-state from rapid intra- and inter-subband thermalization by carrier-carrier and
carrier-phonon scattering. At 5 K, the more uniform ratio of the peak intensities is indicative of frustrated
inter-subband relaxation and direct recombination of excited-state excitons, as discussed in the previous section.

There are no obvious band-gap renormalization (BGR) effects as the QWR carrier density decreases with time.  In the 5 K measurements, a red-shift of 2.5 meV occurs in the position of the $n=2$ peak over the first 400 ps as the exciton population cools. At 60 K this transient shift is not
observed, indicating that a significant number of excitons are thermally activated out of localization sites at this temperature. This is consistent with the
behavior of the shifts observed in the temperature-dependent PL measurements discussed in Sec.\ \ref{pl}, and is also consistent with the difference in Stokes shifts between 5 K and 60 K for the $n=2$ state.

\begin{figure}[tb]
\epsfig{file=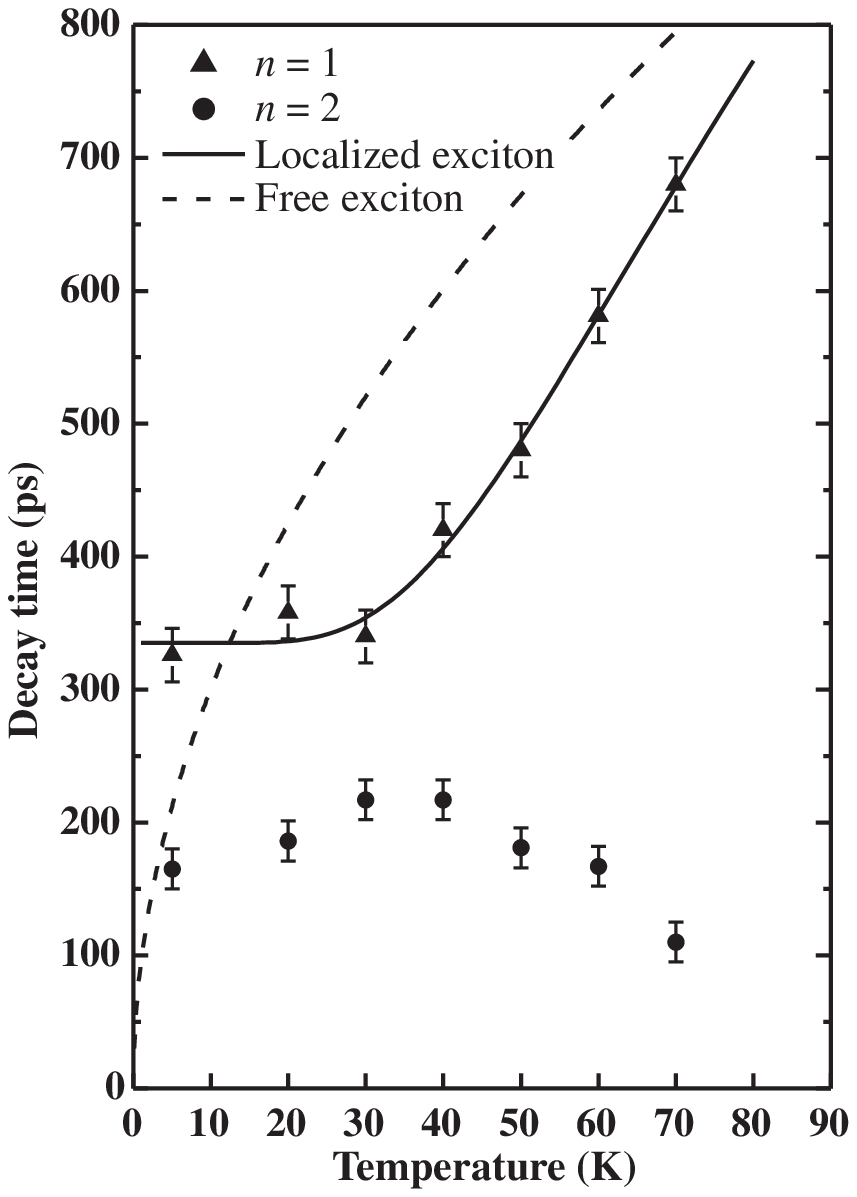,width=8.4cm} \caption{\label{118decayvstemp} Experimental luminescence decay times versus
temperature, for the $n=1$ (triangles) and $n=2$ (circles) QWR states. The dashed line shows the predicted
square-root dependence of the free exciton decay time with temperature, and the solid line includes localization
effects, as given by Eq.\ (\ref{decay}).}
\end{figure}

The luminescence decay times of the QWR states, obtained from Fig.\ \ref{118decay}, are plotted in Fig.\
\ref{118decayvstemp} as a function of temperature. Calculations by Citrin predict that the radiative lifetime
$\tau_{\text{rad}}$ of free excitons in QWRs will  follow a square-root dependence with temperature:\cite{citrin}
$\tau_{\text{rad}} \propto \sqrt{T}$. This assumes that the heterostructure is free of defects, allowing excitons
to propagate freely along the wire axis. After photoexcitation, the excitons are also required to reach a thermal
equilibrium on a time scale short compared to their radiative lifetime. In real heterostructures, the localization
of excitons due to disorder will significantly increase their lifetime over that of free excitons,\cite{citrin2}
resulting in a departure from the square-root dependence at low temperatures. Figure \ref{118decayvstemp} shows
that the decay time of the $n=1$ state is approximately constant at low temperatures, but a rapid increase is
observed above 30 K when phonon detrapping becomes effective.

Citrin's models can be modified for a 1D system of free and localized excitons in thermal equilibrium. By
considering a temperature dependent localized exciton density $N_{\text{loc}}=N_{D}\exp(E_{\text{loc}}/k_{B}T)$,
Lomascolo \textit{et al.}\cite{lomascolo1998} obtain the expression given in Eq.\ (\ref{decay}) for the
temperature dependent PL decay-time $\tau(T)$.
\begin{equation} \label{decay}
\tau(T)=\frac{N_{D}\exp(E_{\text{loc}}/k_{B}T)+\sqrt{\frac{2Mk_{B}}{\hbar^{2}\pi^{2}}}\sqrt{T}}
{\frac{N_{D}}{\tau_{\text{loc}}}\exp(E_{\text{loc}}/k_{B}T)+\frac{1}{\tau_{0}}\sqrt{\frac{2ME_{1}}{\hbar^{2}\pi}}}
\end{equation}
where $N_{D}$ is the effective density of localization centers, $E_{\text{loc}}$ is the exciton localization
energy,  $\tau_{0}$ and $\tau_{\text{loc}}$ are the intrinsic and localized exciton decay times, $M$ is the total
exciton mass, and $E_{1}$ is the maximum kinetic energy of radiative excitons. This method assumes that
nonradiative recombination processes are negligible: we find from our measurements that the integrated intensity
of the $n=1$ peak only starts to decrease at about 100 K indicating that excitonic recombination is predominantly
radiative over the temperature range considered here.

Equation \ref{decay} has been used to fit the experimental points in Fig.\ \ref{118decayvstemp}; $\tau_{0}$ was
fixed while $N_{D}$ and $E_{\text{loc}}$ were adjusted,  and the procedure repeated to give the best fit to the
data (solid line in Fig.\ \ref{118decayvstemp}). Taking $M\simeq 0.3\, m_{0}$ gives $E_{1}=0.09$ meV, and we
obtain values of $N_{D} = 5 \times 10^{4}$ cm$^{-1}$, $E_{\text{loc}} \simeq 11$ meV, $\tau_{\text{loc}}$ = 340 ps and
$\tau_{0}$ = 130 ps, the last being in good agreement with the predicted value.\cite{citrin} It should be noted here that $E_{\text{loc}}$ gives an indication of the energy required to promote localized excitons into quasi-1D continuum states \emph{within the same subband}. Inter-subband relaxation processes will be governed by a separate energy scale due to a laterally varying potential and disorder, as discussed in Sec.\ \ref{plesection}.

Calculations by Oberli \textit{et al.}\cite{oberli1999} give the characteristic exciton localization length
resulting from disorder as $l_{c}\propto \tau_{0}/\tau_{\text{loc}}$. Based on the values obtained above, we
estimate this length to be $l_{c}$ $\simeq$  43 nm, in agreement with Bellessa \textit{et al}.\cite{bellessa} This
expression is valid for $a_{B}\ll l_{c}\ll \lambda_{0}/n$, where $a_{B}$ and $\lambda_{0}/n$ are the 3D exciton
Bohr radius and photon wavelength in the material respectively.

At the $n=2$ level there is a competition between relaxation into the lower subbands, and recombination. The decay
time for the $n=2$ state $\tau_{2}$ is thus a combination of the radiative decay time $\tau_{2,\text{rad}}$  and
the relaxation time $\tau_{2,\text{rel}}$, assuming again that nonradiative recombination is negligible:
$\tau^{-1}_{2}=\tau^{-1}_{2,\text{rad}}+\tau^{-1}_{2,\text{rel}}$. Consequently, it is not straightforward to
extract the two components from the measured decay times. However, the data can be interpreted qualitatively by
assuming that localization effects essentially inhibit all relaxation below 30 K, and that $\tau_{2,\text{rad}}
\simeq \tau_{2}$. This is corroborated by the low-temperature PLE spectra of Fig.\ \ref{118ple1}. Above 30 K, the
increase in the inter-subband relaxation rate is greater than the decrease in the radiative recombination rate,
leading to an overall drop in decay time. The temperature-dependent radiative efficiency $\eta(T)=\tau_{2}/
\tau_{2,\text{rad}}$, can be calculated from the $n=2$ peak intensities in Fig.\ \ref{118tempPL}. By assuming a
similar behavior for $\tau_{2,\text{rad}}$ as is observed for the $n=1$ state, we obtain values of
$\tau_{2,\text{rad}}$ = 425 ps and $\tau_{2,\text{rel}}$ = 280 ps at 60 K. Inter-subband
relaxation by LO-phonon emission is expected to occur on a time scale of $\sim$10$^{-13}$ s for delocalized (1D) excitons, thus
$\tau_{2,\text{rel}}$ gives an indication of the characteristic exciton detrapping time due to LA-phonon scattering.

In contrast to the $n=1$ state, the low temperature radiative lifetime of the $n=2$ exciton is not much greater
than the intrinsic lifetime $\tau_{0}$ obtained above. The oscillator strength for a bound exciton $f$, and hence
the radiative recombination rate, is found to scale as
\begin{equation}
 f\propto\tau^{-1}_{\text{rad}}\propto V_{\text{coh}}(l_{c})/V_{\text{ex}}(l_{c}),\cite{feldmann,bellessa}
\end{equation}
 where $V_{\text{coh}}$ is the coherence volume of the exciton, which represents the spatial
extent of the center-of-mass wave function of the localized exciton, and $V_{\text{ex}}$ is the exciton volume.
$V_{\text{coh}}$ can be taken as the volume of the box in which the exciton is localized and is proportional to
$l_{c}$. For $l_{c} > a_{B}$, the exciton volume is constant and a reduction of $l_{c}$ leads to a reduction of
the  oscillator strength (i.e. an increase in the radiative lifetime above $\tau_{0}$). This is observed for the
$n = 1$ transition and is consistent with the value of $l_{c}$ obtained above ($a_{B} \simeq $ 12 nm for GaAs). If
$l_{c} \alt a_{B}$, and the exciton binding energy is smaller than the localization energy of the carriers, the
electrons and holes are separately confined (strong confinement regime);\cite{kayanuma,iotti} $V_{\text{coh}}$ is
then no longer a valid quantity. The recombination rate will be determined by $V^{-1}_{\text{ex}}$, and will
increase rapidly as $l_{c}$ is reduced due to compression of the electron-hole wavefunction.

The binding energy and radius of the $n = 2$ exciton were determined from magneto-PL measurements to be 9.6 meV and 10 nm
respectively.\cite{jkim,note2} The low temperature $n=2$ Stokes shift implies that the longitudinal (intra-subband) localization energy due to disorder is $\simeq$ 11 meV, which is consistent with the value of $E_{\text{loc}}$ obtained for the $n=1$ state.  This suggests that if the
characteristic localization length of the $n=2$ excitons is less than $\sim$12 nm ($a_{B}$), then these excitons
are strongly confined at low temperatures; hence the reduction in decay time (increased
recombination rate) relative to the $n=1$ exciton.

\section{Conclusion}

In summary, we have studied carrier relaxation processes in an array of v-groove quantum wires. Luminescence
measurements and theoretical models suggest that the pinch-off regions around the QWR have a blocking effect on
real-space relaxation from the SQWs.  In addition, relaxation from higher states of the QWR into the ground state
is inhibited due to disorder induced localization of excitons in the \{311\} facet regions. These effects results
in strong luminescence from the SQW and the first excited state of the QWR below $\sim$50 K, even at low carrier
densities. Evidence for the role of phonons in exciton relaxation has been found in PL, PLE and decay-time measurements; a prominent change in characteristics is observed at $\gtrsim$30 K when fast inter-subband relaxation occurs via phonon-assisted scattering between localized states.
The LA-phonon-assisted exciton detrapping time is estimated to be $\sim$300 ps at 60 K. Values for the low temperature radiative lifetimes of the ground- and first excited-state excitons have been
obtained (340 ps and 160 ps respectively) and interpreted in terms of their corresponding localization lengths.

\section{Acknowledgements}
The authors wish to thank D. Meertens for the TEM image. This work was supported by the EPSRC (UK), and the EC
through the ULTRAFAST network.

\end{document}